# Apports croisés des démarches d'inspection et de test d'usage dans l'évaluation de l'accessibilité de E-services

*Marc-Eric BOBILLIER CHAUMON*
ICTT - ECL
36, avenue Guy de Collongue
BP 163 – F- 69131 Ecully Cedex
marc-eric.bobillier@ec-lyon.fr

*Françoise SANDOZ GUERMOND*
ICTT - INSA
21, avenue Jean Capelle
69621 Villeurbanne Cedex France
françoise.sandoz-guermond@insa-lyon.fr

**RESUME**
Cet article se propose de décrire et de comparer les apports des démarches d'évaluation de l'accessibilité de E-services administratifs effectuées à partir de méthodes d'inspection (ergonomiques et d'accessibilité) et (ii) de tests d'usage. Il ressort que ce sont ces derniers qui présentent le meilleur taux d'identification des problèmes d'usages pour les personnes handicapées.

**MOTS CLES :** Accessibilité, E-services, Personnes handicapées.

**ABSTRACT**
This article proposes to describe and compare the contributions of various techniques of evaluation of the accessibility of E-services carried out starting from (i) methods of inspection (ergonomic and accessibility) and (ii) of tests of use. It show that these are the latter which show the best rate of identification of the problems of uses for the poeple with disabilities

L'administration électronique (E-service) laisse entrevoir de formidables possibilités pour l'amélioration des conditions et de la qualité de vie des personnes handicapées (PH). Elle peut ainsi redonner des opportunités d'action, d'interaction et de décision desquelles ces personnes avaient pu être privées. Cependant, ces nouveaux services sont trop souvent conçus sans tenir compte des caractéristiques de cette population à besoins spécifiques ; ce qui a pour conséquence d'accentuer davantage encore l'exclusion dont elles sont victimes.

L'objectif de cette communication est de confronter des méthodes d'évaluation ergonomique et d'accessibilité afin d'une part, d'identifier les difficultés d'usage des PH avec les E-services, et, d'autre part, de repérer les apports et limites de ces différentes approches. Nous serons ainsi amenés à explorer le concept d'accessibilité technologique puis à présenter la démarche appliquée et les principaux résultats de nos analyses.

## APPROCHE THEORIQUE DE L'ACCESSIBILITE

Le WAI [17] (Initiative pour l'Accessibilité du Web lancé en 1997 par le W3C) définit l'accessibilité par le fait que des personnes puissent "*percevoir, comprendre, naviguer et interagir de manière efficace avec l'internet, mais aussi créer du contenu et apporter leur contribution au Web*". Bien que l'ergonomie intègre cette dimension dans sa réflexion (dans l'idée d'adapter les dispositifs aux spécificités et aux besoins des usagers ainsi qu'aux caractéristiques de leur activité), il est possible néanmoins, comme le propose [15] de distinguer ces deux approches : "*si l'ergonomie se base sur une approche plus globale qui résiste généralement aux tentatives de spécifications*", l'accessibilité "*se base [quand à elle] sur des spécifications techniques précises issues de règles*" (p 72) .

Ainsi, le WAI propose 14 directives pour évaluer la conformité des sites. Celles-ci sont d'ailleurs reprises par des projets de loi sur l'accessibilité des sites administratifs : section 508 aux USA, loi sur l'accessibilité numérique des administrations en France. Ces principes donnent également lieu à des méthodes d'évaluation et de labellisation comme Blindsurfeur en Belgique, See it Right en Angletterre [13]. En France, Braillenet [5] propose un label "Accessiweb" qui comporte des recommandations générales et des points de contrôle pour définir trois niveaux de qualité d'un site (bronze, argent et or). Mais parce que la mise en œuvre de ces check-lists est souvent longue et fastidieuse, et donc lente et chère, des outils d'inspection automatisée (comme Bobby, Infocus, A-Prompt…) sont utilisés pour faciliter l'identification des problèmes d'accessibilité [12]. Dans ce même souci d'efficacité, [8] a développé neuf heuristiques d'évaluation de l'accessibilité afin de dépister les principales barrières technologiques aux PH. Enfin, des méthodes ergonomiques plus "classiques" proposent aussi la prise en compte de cette dimension, comme le propose [1] avec, parmi les huit critères listés, ceux de l'adaptabilité et de la compatibilité.

Pourtant, malgré ces diverses incitations et méthodes, les études menées montrent toutes la très faible accessibilité des E-services, quelque soit l'administration ou le pays concerné [2][4][9] : plus de 75 % présenteraient ainsi des défauts d'accessibilité de niveau 1 (des points de contrôle WAI), c'est-à-dire rendant l'accès impossible à des PH. Plusieurs raisons à cela : l'absence de formation en accessibilité des concepteurs, des coûts de développement *a priori* trop importants, le manque d'intérêt pour ces profils d'usagers, la crainte d'appauvrir le site [14]. De plus, comme l'indique **[7]**, la recherche d'utilisabilité (par des composants graphiques : icônes, menu déroulant, onglets…) poserait des problèmes significatifs d'accessibilité. Autrement dit, certaines innovations ergonomiques iraient à l'encontre de l'accessibilité souhaitée.

Dans ces conditions, et pour assurer un niveau acceptable d'utilisabilité pour le plus grand nombre, le croisement des évaluations -ergonomique et d'accessibilité- complété par des études d'usage semblent être une solution de compromis pour appréhender plus globalement les problèmes d'usage qui peuvent se poser à l'ensemble des utilisateurs ; comme l'ont d'ailleurs déjà montré [3] pour des diagnostics plus classiques de sites WEB ou [16] pour l'évaluation de l'accessibilité de sites administrifs..

## METHODE MISE EN OEUVRE

Pour apprécier la qualité des sites administratifs et confronter les apports et limites de chaque méthode, nous nous sommes livrés à deux séries d'analyse sur deux sites : l'ANPE et la mairie de Vandoeuvre :

a) D'abord une approche par inspection (ergonomique et d'accessibilité) réalisée par deux groupes de quatre experts afin de repérer les principaux écueils des sites. La grille ergonomique a été élaborée à partir de 75 critères ventilés dans huit rubriques (celles de Bastien et Scapin [1]) que nous détaillerons plus bas.

Celle d'accessibilité a été conçue à partir de la grille Accessiweb (Braillenet (partenaire de cette étude) qui propose 55 critères pour valider le niveau bronze d'accessibilité [5]. Ces critères ont été répartis par les experts dans les huit même rubriques de l'inspection ergonomique afin de disposer de trame d'évaluation comparable. Chaque groupe devait ainsi évaluer la pertinence d'une affirmation (check-list) sur une échelle allant de 1 (pas du tout d'accord/satisfaisant) à 4 (tout à fait d'accord/très satisfaisant) et justifier leur choix. Par exemple "*il y a toujours possibilité d'annuler ses actions*". Une réunion de concertation finale permettait d'harmoniser les notes de chaque groupe.

b) Des tests d'usage ont été réalisés sur 10 sujets aveugles qui présentaient des caractéristiques socio-biographiques équivalentes (âge, sexe, formation, activité…), seul leur maîtrise des environnements Web les distinguaient (5 experts *vs* 5 novices).

Les scénarii à réaliser portaient sur trois types de E-service: les services informationnels (*rechercher une information administrative*) ; interactionnels (*participer à un forum*) et transactionnels (*remplir une demande d'acte de naissance*). Par des techniques de verbalisations simultanées et d'observation, nous avons relevé les indicateurs portant sur les difficultés d'usage.

## PRINCIPAUX RESULTATS

### Principaux résultats des inspections

Les évaluations effectuées par les deux groupes d'experts sont résumées dans le tableau 1. Il confronte les résultats obtenus avec les deux méthodes d'inspection (accessibilité/ergonomique) sur la base des 8 rubriques : guidage (*a*), charge de travail (*b*), contrôle explicite (*c*), adaptabilité (*d*), gestion des erreurs (*e*), homogénéité/cohérence (*f*), signifiance des codes (*g*), compatibilité (*h*)

| | | Critères ergonomiques considérés | | | | | | | | |
|---|---|---|---|---|---|---|---|---|---|---|
| | | A | B | C | D | E | F | G | H | Moy |
| Site ANPE | Insp. Acce | 2.51 | 3.90 | 2.18 | 2.68 | 3.33 | 3.32 | 1.17 | 4 | 2.89 |
| | Insp. Ergo | 3.42 | 3.64 | 4 | 3.50 | 2.81 | 3.80 | 3.83 | 4 | 3.63 |
| Site Mairie | Insp. Acce | 2.96 | 3.13 | 2.13 | 3.80 | 1.92 | 2.87 | 2.03 | 4 | 2.69 |
| | Insp. Ergo | 2.70 | 2.93 | 4 | 3 | 1.80 | 2.60 | 3.60 | 3.80 | 3.05 |

***Tableau 1 :*** *Tableau des évaluations obtenues par l'inspection ergonomique classique et d'accessibilité pour les deux sites*

D'une manière générale, il ressort que :
- Le groupe 1 (inspection classique) a globalement mieux noté les sites que le groupe 2 (accessibilité).
- le site de l'ANPE obtient des évaluations légèrement meilleures à celles de la mairie, et ce, quelque soit la méthode d'inspection mise en œuvre. Ce qui est assez surprenant puisque la mairie de Vandoeuvre bénéficie du niveau bronze du label Accessiweb.
- Il existe des critères convergents entre les deux méthodes d'inspection : ce sont les rubriques "charge de travail" (*b*), "Homogénéité-cohérence" (*f*) et "compatibilité" (*h*). A titre d'exemple, les deux groupes d'experts pointent sur le non-respect du principe d'homogénéité (site mairie), notamment à cause de la barre de navigation qui se réorganise dynamiquement selon les choix de l'utilisateur ou encore de la charge de travail induite par la densité des informations à percevoir et à mémoriser (de 18 à 30 sous-rubriques sur le site mairie).
- On relève des divergences entre les évaluations, en particulier sur le "contrôle explicite" (*c*), "l'adaptabilité" *(d)*, "gestion des erreurs" *(e)* et "signifiance des codes et dénominations" *(g)*. Les notes attribuées par les experts de l'accessibilité sont en général plus basses que celles fournies par les experts de l'ergonomie classique. Ainsi, selon l'inspection ergonomique du critère "signifiance des codes" (g), les éléments textuels et iconographiques des deux sites paraissent intuitifs et compréhensibles (3,83 et 3,60). Alors que l'approche par l'accessibilité pointe sur le manque de précision et d'explicitation des commentaires alternatifs associés aux boutons de validation des formulaires (1,17 et 2,03) : quelque soit le type de champ de saisie à valider, l'alternative textuelle "OK" est systématiquement accolée au bouton éponyme.

### Principaux résultats des tests utilisateurs

Les tests utilisateurs ont permis : (a) de valider, préciser, et/ou de compléter certains résultats des inspections menées, (b) de découvrir des problèmes d'usage non détectés par les analyses, ou encore (c) de contredire certaines évaluations des experts.

**a)** ***Critères confirmés par les tests d'usage*** : on a pu observer que la densité d'information -relevée par le critère charge de travail- induit effectivement une charge cognitive importante qui contrarie la navigation et l'implication de l'usager aveugle dans le site. Il doit en effet à la fois écouter le synthétiseur vocal, mémoriser les rubriques, savoir se positionner dans l'arborescence du site tout en développant des stratégies pour atteindre l'objectif fixé par le scénario : "*Vu qu'on n'a pas une vue d'ensemble sur la page, on est obligé de tout parcourir, on est obligé d'être intuitif... de faire des hypothèses sur les résultat d'un lien. Il faut se demander comment le concepteur aurait pu nommer le lien*".

Autre critère validé par les tests d'usage, le fait que a réorganisation erratique de la barre de navigation déstabilise les usagers handicapés (Critère homogénéité). Ils n'ont en effet ni la capacité perceptive, ni les moyens techniques pour détecter ce changement. Du coup, cela rend caduque la représentation mentale qu'ils s'étaient construites de l'arborescence du menu principal et vont redoubler d'effort pour retrouver les rubriques qui ont bougé.

*b) Nouveaux problèmes d'usage révélés par les tests*
Parmi les obstacles non détectés par l'inspection, on peut citer le problème de remplissage des formulaires qui oblige l'usager à passer successivement de la lecture des libellés (par exemple, "saisir votre nom") à la saisie de données. Cette alternance n'est pas géré automatiquement par le lecteur d'écran Jaws mais doit être déclenché manuellement par l'usager à l'aide d'une commande spécifique (quand il y pense). Du coup, il se retrouve souvent à lire des champs de saisie (où il n'y a rien) et à vouloir entrer de l'information dans les libellés ! Finalement, les usagers dépensent beaucoup de temps et d'énergie à comprendre et à récupérer les erreurs de saisie. Ce qui accroît là aussi la charge cognitive.

*c) Critères "contredits" par les tests d'usage :* Pour le critère "charge de travail", l'affichage de sous-menus contextuels dans la barre de navigation du site de la mairie avait été noté plutôt positivement par les experts puisqu'il permettait de limiter la densité informationnelle. Or, les tests d'usage montre que les usagers ne perçoivent pas ces sous-menus car ils vont sauter les barres afin de ne pas lire ces informations récurrentes.

De même, le critère "guidage" relevait favorablement la présence d'un exemple dans les champs de saisie pour aider l'utilisateur à entrer les données. Or, les tests montrent que ce type d'incitation génère un grand nombre d'erreurs, notamment pour le moteur de Recherche de la Mairie. : la donnée saisie est en effet automatiquement concaténée avec l'information préexistante (Ici "Rechercher"). La présence de ce mot est donc en soit une incitation pour des personnes "valides" mais sa persistance déclenche des erreurs que les personnes aveugles ne peuvent détecter.

**DISCUSSION- CONCLUSION**

Ces analyses nous conduisent à nous interroger sur la portée de chaque méthode d'évaluation : le fait de considérer un site accessible induit-il le fait que tous les critères ergonomiques classiques soient respectés ? Et inversement, le respect des critères ergonomiques classiques implique-t-il le respect des règles d'accessibilité (via les critères adaptabilité et compatibilité notamment) ?

| Critè-res | Principaux problèmes identifiés | Inspect classique | Inspect Access | Tests util. |
|---|---|---|---|---|
| Guidage | Libellé peu explicite, polysémie des termes | | | √ |
| | Identification des cellules d'un tableau | | √ | √ |
| | Structuration linéaire des pages sous forme de tableau | | √ | √ |
| | Alternatives textuelles absentes ou peu explicites | | √ | √ |
| | Liens (identification, changement de couleurs…) | √ | | |
| Contrôle explicite | Ouverture intempestive de fenêtres | √ | √ | √ |
| | Affichage des sous menus contextuels non signalé | | | √ |
| Charge de travail | Densité, identification et répartition inadaptées des informations sur les pages | √ | √ | √ |
| | Recours à des ascenseurs verticaux | √ | | |
| | Non désactivation du lien par rapport à la page visitée | √ | √ | √ |
| Gestion des erreurs | Qualité des messages d'erreurs | √ | √ | √ |
| | Saisie erronée non détectée | √ | | √ |
| | Champ obligatoire non indiqué | √ | | √ |
| | Persistance de données dans les champs de recherche | | | √ |
| Homogénéité | Changement erratique des barres de navigation | √ | √ | √ |
| Adaptabilité | Absence d'alternative au java script | | √ | √ |
| | Alterner manuellement "lecture du libellé/saisie dans le champ" | | | √ |
| Nombre de problèmes repérés | | 8 | 9 | 15 |

*Tableau 2:* Tableau récapitulatif des problèmes identifiés par chaque méthode d'évaluation

Les résultats obtenus sur les deux sites tendraient à montrer un taux de recouvrement partiel entre les deux méthodes d'évaluation. Il serait donc préférable de se livrer à une inspection par l'intermédiaire de ces deux approches pour s'assurer d'un diagnostic optimal plutôt que d'en privilégier une seule. Ce que souligne d'ailleurs [13] qui remarque que si la conception des technologies cherche à favoriser l'accessibilité pour les PH, ce sont surtout les principes de l'ergonomie classique qui sont le plus souvent employés par les concepteurs au détriment des principes d'accessibilité. L'approche classique conçoit les systèmes en laissant penser que les utilisateurs ont des buts spécifiques -que l'environnement doit satisfaire- alors que les principes d'accessibilité prennent en compte les limites et les difficultés physiques, perceptives et mentales des usagers handicapés. Pour ces raisons, les principes de l'ergonomie classique ne sauraient remplacer les règles d'accessibilité dans la conception, bien que ces deux approches soient complémentaires. Outre l'exhaustivité assurée, cette double approche permet également de se prémunir contre certains biais de la conception universelle [6], notamment lorsque des améliorations à destination de certaines catégories d'usagers à besoins spécifiques peuvent se révéler pénalisantes pour d'autres groupes d'utilisateurs, handicapés ou non, et inversement (*cas des sous-menus contextuels par exemple*).

Pour ces raisons, les tests utilisateurs semblent offrir le meilleur compromis comme l'indique le tableau 2 qui compare la "performance" de chaque approche dans l'identification des problèmes d'usage. Il ressort que ce sont les tests qui permettent de retrouver la quasi-totalité des défauts identifiés par les deux autres méthodes d'inspection (15/17) ; en en validant la plupart et en en découvrant aussi de nouveaux non détectés par les démarches traditionnelles d'évaluation (5/17). Ceux qui n'ont pas été identifiés par ces tests ne représentaient cependant pas de barrières rédhibitoires pour les sujets aveugles (*utilisation des ascenseurs…*), même s'ils peuvent néanmoins poser certains problèmes au cours de l'interaction (*changement de couleurs des liens non signalés…*).

Cette nécessaire complémentarité entre ces deux méthodes d'évaluation (inspection et tests d'usage) avait également été souligné par Jeffris (1991) (cité par [3]) qui montrait dans son étude que les évaluations par experts avaient manqué à peu près la moitié des problèmes soulevés par les tests d'utilisabilité et inversement. [3] précisant que chaque approche permet de découvrir des problèmes de type différent : les tests d'utilisabilité trouvant des problèmes reliés à une tâche précise (celle établie par les scénarios construits) tandis que les évaluations heuristiques (ou par inspection) trouvent des problèmes plus généraux. Le choix de la technique d'évaluation dépend donc des buts de l'évaluation, du genre de problèmes recherché et des ressources disponibles. En tout état de cause, il s'agit de les mener de manière parallèle et complémentaire pour confronter des analyses soit très standardisées mais qui paraissent au final assez "désincarnées" (cas des inspections), soit proches d'une certaine réalité d'usage, mais échappant à toute systématisation (cas des tests).

**REMERCIEMENTS**

**BIBLIOGRAPHIE**

[1] Bastien, J.M.C., Leulier, C., & Scapin, D.L. L'ergonomie des sites web. In J.-C. Le Moal & B. Hidoine (Eds.), *Créer et maintenir un service Web,* ADBS, Paris, 1998, pp. 111-173.

[2] Beckett D.J. 30% accessible - a survey of the UK Wide Web : Computer Networks and ISDN Systems, Web 29, 1997, pp. 1367-1375

[3] Boutin M., Martial O. *Évaluation de l'utilisabilité d'un site Web : tests d'utilisabilité versus évaluation heuristique.* Mai 2001. Rapport technique CRIM Québec

[4] Braillenet. *Pour une meilleure accessibilité des sites publics aux personnes handicapées*. 2002, Disponible adresse : http://www.dusa.gouv.fr/IMG/pdf/braillenet.pdf

[5] Braillenet-accessiweb *Certifier l'accessibilité des sites Web*, 2005, disponible à l'adresse http://www.accessiweb.org/

[6] Brangier, E., Barcenilla, J. *Concevoir un produit facile à utiliser : adapter les technologies à l'homme,* , Editions d'organisation, Paris, 2003.

[7] Chiang M. Cole R., Gupta S., Kaiser G., and Starren, J. *Computer and World Wide Web Accessibility by Visually Disabled Patients: Problems and Solutions.* Survey of ophtalmology, Vol 50, No 4, 2005, pp 394-405.

[8] Englefield, P.J. HEDB: a software tool to support heuristic evaluation. In: Gray, P., et al. (Eds.), *Proceedings of HCI Designing for Society*, Research Press International, 2003, pp 507-521

[9] Fagan J.C. and Fagan B. *An accessibility study of state legislative Web sites Government Information Quarterly* 21, 2004, pp. 65–85

[10] Fiset J.Y, *Services électroniques aux citoyens et aux entreprises. Étude sur l'accessibilité*, no 2002-115- .9, 4 mars 2003, produit pour le CEFRIO par Systèmes Humains-Machines Inc

[11] Foley, P., Ximena A., Fisher J., Bradbrook G.,. *eGovernment : Reaching socially excluded groups?*, 2005, disponible à l'adresse http://www.idea-knowledge.gov.uk /idk/aio/1075006

[12] Ivory, M., Mankoff, J., & Le, A. *Using automated tools to improve web site usage by users with diverse abilities.* IT and Society, Vol. 1, No. 3, 2003, pp. 195–236

[13] Jaeger P.T. *Telecommunications policy and individuals with disabilities: Issues of accessibility and social inclusion in the policy and research agenda.* Telecommunications Policy, Vol 30, 2006, pp.112–124

[14] Lazar J., Dudley-Sponaugle A., Greenidge K.D (2004) *Improving web accessibility: a study of webmaster perceptions,* Computers in Human Behavior 20 269–288

[15] Richards J.T., Hanson V.L., *Web accessibility : a broader view*, May 2004, Proceedings of the 13th international conference on World Wide Web, pp .

[16] Theofanos T., Redish J. T (2003) *Bridging the gap: between accessibility and usability*, interactions, Volume 10 Issue 6, pp 37-51.

[17] W3C-WAI Introduction à l'accessibilité du Web. 2005. Disp. à l'adresse : http://w3qc.org/docs/ accessibilite.html